\documentclass[aps,pra,twocolumn,a4paper,amsmath,amssymb,showpacs,%
superscriptaddress,floatfix,]{revtex4}

\usepackage{color}
\usepackage{graphicx}
\usepackage{epstopdf}

\newcommand{\Op}[1]{\boldsymbol{\mathsf{\hat{#1}}}}

\newcommand{\Fkt}[1]{\,\mathsf {#1}}

\def\openone{\leavevmode\hbox{\small1\kern-3.3pt\normalsize1}}

\begin{document}

\title{Correlation dynamics after short-pulse photoassociation}

\author{Christiane P. Koch}
\email{ckoch@physik.fu-berlin.de}
\affiliation{Institut f\"ur Theoretische Physik,
  Freie Universit\"at Berlin,
  Arnimallee 14, 14195 Berlin, Germany}
\author{Ronnie Kosloff}
\affiliation{Institute of Chemistry and
  The Fritz Haber Research Center, 
  The Hebrew University, Jerusalem 91904, Israel}

\date{\today}

\begin{abstract}
Two atoms in an ultracold gas are correlated at short inter-atomic
distances due to threshold effects where the potential energy of their
interaction dominates the kinetic energy. The correlations manifest
themselves in a distinct nodal structure of the density matrix at short
inter-atomic distances. Pump-probe spectroscopy has recently been
suggested [Phys. Rev. Lett. 103, 260401 (2009)] to probe these pair
correlations: A suitably chosen, short photoassociation laser pulse 
depletes the ground state pair density within the photoassociation
window, creating a non-stationary wave
packet in the electronic ground state. The dynamics of this
non-stationary wave packet is monitored by time-delayed probe and
ionization pulses. Here, we discuss how the choice of the
pulse parameters affects experimental feasibility of this pump-probe
spectroscopy of two-body correlations. 
\end{abstract}

\maketitle


\section{Introduction}
\label{sec:intro}

The simplest description of Bose-Einstein condensation (BEC) 
considers an ensemble of
indistinguishable particles all in the same quantum state
\cite{PethickSmith,LeggettRMP01}. Such a
picture leads to a mean-field representation of the many-body
wave function as a direct product 
of the single-particle wave functions. This
framework has been successful
in describing many aspects of BEC physics. 
Mean-field approaches neglect  two-body correlations which arise due
to the long range interaction between two particles. 
In a microscopic theory,  pair correlations can be incorporated
explicitly by expanding the correlation functions  into cumulants
\cite{KoehlerPRA02,KoehlerPRA03}. Alternatively, one can also work
directly with the  correlation functions \cite{PascalPRA03,PascalPRA06}.  
Can one envision directly  measurable consequences of the pair 
correlations? We have recently answered this question in the
affirmative~\cite{MyPRL09} combining our previous work on short-pulse
photoassociation~\cite{JiriPRA00,ElianePRA04,MyPRA06a,MyPRA06b,MyPRA04,MyPRA08}
with a treatment of many-body pair 
correlations~\cite{KoehlerPRA02,KoehlerPRA03,PascalPRA03,PascalPRA06}. 
Here we discuss in detail how the choice
of experimentally tunable parameters determines feasibility of the
proposed experiment.

\begin{figure}[b]
  \centering
  \includegraphics[width=0.95\linewidth]{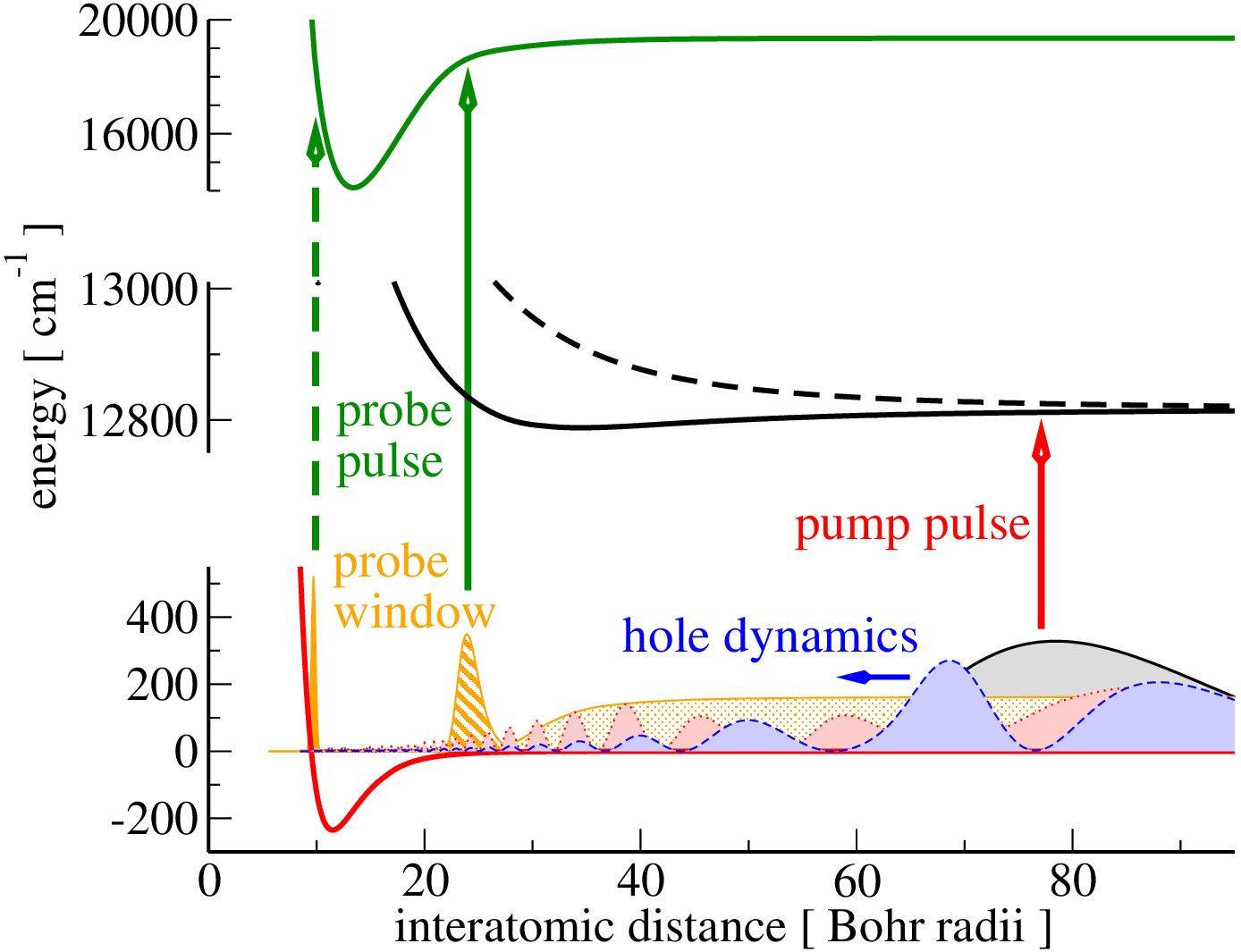}
  \caption{(Color online)
    Pump-probe spectroscopy of the dynamical hole:
    A photoassociation pump pulse excites population from the ground
    state to the first excited state, leaving a hole in the ground
    state wave function and transferring momentum to it. 
    The ground state wave function (bottom panel)
    is depicted before the pulse (black
    solid line),
    at $t_{max}+24\,$ps, i.e. just after the pulse (blue dashed line)
    and at $t_{max}+126\,$ps (red dotted line). For red
    detuned pump pulses, the hole is accelerated toward shorter
    distances which leads to an enhancement of population at short
    distances. 
    A suitably time-delayed REMPI probe pulse can measure this
    population enhancement. 
    The effect of the probe pulse is modeled
    by a window operator (shown in thin orange lines) where different
    fillings reflect 
    different probe central frequencies. Another option is to
    choose the same central frequencies for pump and probe, requiring
    an additional REMPI pulse, cf. Fig.~1 of Ref.~\cite{MyPRL09}.
  }
  \label{fig:scheme}
\end{figure}
Photoassociation assembles two atoms to a molecule with laser
light~\cite{JonesRMP06}.
Considerable experimental effort has already been devoted to
photoassociation with short laser pulses. 
Early experiments were carried out at room
temperature in mercury vapor \cite{DantusCPL95}. In the ultracold
regime, first experiments aimed at femtosecond photoassociation of
rubidium dimers
have led to dissociation of molecules created by the trap lasers rather
than association \cite{SalzmannPRA06,BrownPRL06}.  Recently, evidence of
coherent formation of molecules in the electronically excited state
was provided in a pump-probe experiment with chirped
femtosecond laser
pulses~\cite{SalzmannPRL08,WeiseJPB09,MullinsPRA09,MerliMyPRA09,McCabePRA09}.  
The proven experimental ability of pulsed photoassociation paves the
way for developing a pump-probe spectroscopy to study ultracold two-atom
correlation dynamics~\cite{MyPRL09}. 

Photoassociation takes place at the same inter-atomic  
distances where pair correlations are significant. At ultralow
temperatures, photoassociation spectroscopy with continuous-wave (cw)
lasers is sensitive to the position of the nodes of the scattering 
wave function describing two colliding atoms
\cite{FrancoiseReview}. The two-atom scattering wave function is
closely related to the 
reduced pair wave function characterizing the two-body
correlations~\cite{PascalPRA03,MyPRL09}. 
The observed modulation  in the
spectroscopic features therefore serves as a probe of the static inter-atomic
correlations. 
If one is to measure two-body correlations \textit{dynamically},
a non-stationary initial state has to be generated.
This can be achieved by interaction with an external
field which is fast
relative to the ensuing dynamics.  Therefore the short-pulse photoassociation
scenario~\cite{JiriPRA00,ElianePRA04} can be employed to  
generate a non-stationary initial state by a pump pulse and then follow
its dynamics by a probe pulse~\cite{MyPRL09}. 
The idea, sketched in Fig.~\ref{fig:scheme},
is based on impulsive excitation where a selected part of the
ground state probability density is suddenly removed to the excited
potential energy surface \cite{GuyI3-97}.
As a result the ground state phase space
density  is no longer stationary. 
The void in the ground state probability density is termed the
dynamical 'hole'~\cite{ElianePRA07,MyPRL09}. The impulsive excitation
by the pump pulse also 
transfers momentum to the ground state wave function.   The 'hole'
dynamics can be monitored by a suitably time-delayed probe pulse,
detecting the enhancement of ground state density at a specific
location by resonantly
enhanced multi-photon ionization (REMPI), cf. Fig.~\ref{fig:scheme}, or
by a combination of probe and ionization pulse~\cite{MyPRL09}.

As in any stroboscopic measurement, 
the pump and probe pulses are required to act on
a timescale  shorter than the observed dynamics \cite{banin94,GuyI3-97}.
The correlations in  ultracold atomic gases are due to  threshold
effects, where the kinetic energy of the colliding atoms is smaller or
equal to the potential energy.
The timescale of the correlations is related to the period of the last bound level, 
$T_{vib} \approx 2 \pi \hbar \frac{ \partial v}{\partial E}$,
or to the scattering length $a_S$, $ \tau \approx \mu a_S / \hbar k$ 
\cite{ElianePRA04}. For the example of $^{87}$Rb with a scattering
length of $\approx~ 100\,a_0$, the correlation timescale is typically of
the order of 100$\,$ps to 1000$\,$ps. 
This timescale defines an upper limit to the pulse duration. 
A more quantitative estimate of this limit will be derived below by
comparing the change in kinetic energy to the spectral bandwidth of
the pulse.  
On the other hand, the pulse duration cannot be too short since then
the spectral bandwidth would be very broad. This would result in the
dynamical 'hole' being smeared over a large range of distances,
hampering precise control of the 'hole' location. 
Therefore the optimal pump pulse is a compromise
between impulsive excitation and sharp spectroscopic features. 

The insight gained in the photoassociation
studies~\cite{JiriPRA00,ElianePRA04,MyPRA06a,MyPRA06b,MyPRA04,MyPRA08}
can be used to determine the pulse parameters when
constructing a
spectroscopic probe of the two-atom correlation dynamics~\cite{MyPRL09}.
The choice of the pulse parameters controls the location
and the momentum of the dynamical 'hole' \cite{MyPRA06b,ShimshonPRA07,MyPRL09}. 
It can be analyzed by projecting onto the stationary energy eigenfunctions.
From this viewpoint, the pump pulse redistributes 
the initial scattering state 
to bound molecular levels and other scattering wave functions. 
The 'hole' location is determined by the detuning from the atomic
resonance \cite{MyPRL09,MyPRA06b}.
The momentum of the 'hole' can be controlled by proper chirp parameters
\cite{ShimshonPRA07,ElianePRA07} and by the choice of an attractive or
repulsive potential of the electronically excited state
\cite{ShimshonPRA08}, i.e. by the pulse detuning.

The paper is organized as follows. Section~\ref{sec:model} introduces
the theoretical description of the excitation by the pump pulse and 
'hole' dynamics in a two-state picture as well as the model of the probe
pulse absorption by a window operator.  The effect of the pump pulse
is studied in Section~\ref{sec:pump}, and the observation of the
'hole' dynamics is detailed in Section~\ref{sec:probe}. 
Section~\ref{sec:concl} concludes. 

\section{Model}
\label{sec:model}

We consider two scattering atoms interacting via the electronic
ground potential. The center of mass motion is integrated analytically and 
omitted from this description.  For 
sufficiently low temperatures only $s$-wave encounters contribute.
The ground surface supports both bound levels and
scattering states. 
A stationary initial  state is considered. For a BEC this state can be well 
approximated by a scattering state \cite{PascalPRA03,PascalPRA06} with
momentum determined by 
the chemical potential. For a thermal state the initial state is 
a thermal mixture of scattering states \cite{MyJPhysB06}.
In both cases the two body correlation decay at long interatomic
distance due to the vanishing interaction. This means that for both
cases the simulation can be based on an initial scattering state 
with a finite cutoff in internuclear distance.
A photoassociation laser couples the electronic ground state to an
electronically excited state where the potential can be either
repulsive or attractive, cf. Fig.~\ref{fig:scheme}.

\subsection{Excitation by a pump pulse and dynamics}
\label{subsec:PA}

The time-dependent Schr\"odinger equation is solved for two electronic
states, 
\[
  i\hbar\frac{\partial}{\partial t} |\psi(t)\rangle =
  \Op{H} |\psi(t)\rangle \quad  \mathrm{and} \quad 
  \langle R |\psi(t)\rangle =
  \begin{pmatrix} \psi_\mathrm{g}(R;t) \\ \psi_e(R;t)  \end{pmatrix}
\]
with  the Chebychev propagator \cite{RonnieReview88}.  
The Hamiltonian in the rotating-wave approximation reads
\begin{equation}
  \label{eq:H}
      \Op{H} =
    \begin{pmatrix}
      \Op{T} + V_{g}(\Op{R}) & \Op{\mu} \, E_0 S(t) \\[2ex]
      \Op{\mu} \, E_0 S^*(t) &
      \Op{T} + V_\mathrm{exc}(\Op{R}) - \hbar\Delta_\mathrm{L}
    \end{pmatrix} \, ,
\end{equation}
where $\Op{T}$ denotes the kinetic energy,  $V_{g/exc}$ the ground and
excited state potential energy curves, and $\Op{\mu}$ the transition
dipole operator. 
Since the excitation occurs at long range, the $R$-dependence of
$\Op{\mu}$ can be neglected. 

The pulse parameters appearing
in  Eq.~(\ref{eq:H}) are $E_0$ corresponding to the
maximum field amplitude, a slowly varying shape of the laser
pulse $S(t)$, taken to be Gaussian, and the detuning $\Delta_\mathrm{L}$. 
The latter is given in terms of the atomic
resonance line and the central laser frequency, $\Delta_\mathrm{L} =
\omega_{at} - \hbar\omega_L$.
The pulse energy is calculated from the maximum field amplitude and
transform-limited full-width half-maximum (FWHM) $\tau$, 
\[
\mathcal{E} = \epsilon_0 c \sqrt{\pi} \pi \frac{ r^2 E_0^2 \tau}{2\ln 2}
\]
with $\epsilon_0$ the electric constant, $c$ the speed of light and $r$ the
radius of the laser beam ($r=100\mu$m is assumed throughout this work).

For blue detuning ($\Delta_\mathrm{L} > 0$) excitation into
the upper-most repulsive potential correlating to the $5s+5p_{3/2}$ asymptote, the 
$0_\mathrm{g}^+(5s+5p_{3/2})$ state, is assumed. 
For red  detuning ($\Delta_\mathrm{L} < 0$), $V_\mathrm{exc}$ is taken to be the 
$0_\mathrm{g}^-(5s+5p_{3/2})$ state, well-known for its purely long-range
attractive well. 
In both cases, $V_\mathrm{g}$ corresponds to the lowest triplet state,
$a^3\Sigma_u^+(5s+5s)$. The potentials at short range are found in 
\cite{AymarJCP05,LozeilleEPJD06}. They are connected to analytical long-range
dispersion potentials 
$ C_3/\Op{R}^3 + C_6/\Op{R}^6+C_8/\Op{R}^8$ with coefficients taken
from \cite{MartePRL02,GuterresPRA02}.
We restrict our model to this two-state description, considering only the
lowest triplet state as electronic ground state. In principle,
a laser pulse excites scattering amplitude from both the lowest
triplet state and the singlet ground state into all states correlating
with the $5s+5p_{3/2}$ asymptote. The resonance condition
might differ for singlet and triplet, giving rise to different probe
signals for singlet and triplet. As already pointed out in
Ref.~\cite{MyPRL09}, this effect may be used to map out the pair
correlation functions for the singlet and triplet components.

The Hamiltonian is represented on a Fourier grid with variable step
size \cite{SlavaJCP99,WillnerJCP04,ShimshonCPL06}. Our  grid
extends to about $R_\mathrm{max}=28000\,$a$_0$. This ensures that
no reflection of the wave packet at the outer boundary of the grid occurs:
Within 10$\,$ns, the fastest wave packet components occurring in our
simulations reach about $R=800\,$a$_0$. 
A single scattering state with a scattering energy corresponding to
20$\,\mu$K is considered as initial state in the calculations
below. The effects of averaging over many thermally populated
scattering states and the influence of 
temperature have been discussed in our previous work \cite{MyPRL09}. 
Here we focus on how the pulse parameters need to be chosen in order
to realize a pump-probe spectroscopy that yields a sufficiently high
signal to be experimentally feasible.

\subsection{Modelling the absorption of the probe pulse by a window operator}
\label{subsec:window}

In order to avoid a separate calculation for each pump-probe time
delay, the total absorption is 
represented by a window operator $\Op{W}(\Op{R})$~\cite{cina97,Jiri00},
\[
  \Delta E = -\hbar \omega_L \Delta N_\mathrm{g} 
  \approx  - \hbar \omega_L \langle \psi_\mathrm{g}(t) | \Op{W}(\Op{R}) |
  \psi_\mathrm{g}(t)\rangle\,, 
\]
with 
\begin{equation}
  \label{eq:window}
  \Op{W}(\Op{R}) =  \pi (\tau_p E_{p,0})^2
  \Fkt{e}^{-2 \Op{\Delta}^2 (\Op{R})\tau_p ^2}\cdot 
  \Op{\mu}^2_p (\Op{R})\,.
\end{equation}
The window operator contains 
the probe pulse parameters FWHM $\tau_p$ and maximum field amplitude
$E_{p,0}$. The probe pulse central frequency determines the difference
potential, 
\begin{equation}
\Op{\Delta}(\Op{R}) = V_{p}(\Op{R}) - V_\mathrm{g}(\Op{R}) - \hbar \omega_p 
\label{eq:diff}
\end{equation}
with $V_{p}(\Op{R})$ denoting the potential which is accessed by the
probe. The $R$-dependence of the transition dipole moment $\Op{\mu}_p$
is neglected.  The physical concept
of the window operator is to collapse the observation process which is
completed in a time proportional to the probe pulse duration $\tau_p$ to a single
instant in time $t_p$. The finite width in time which corresponds to a finite
width in frequency $\Delta \omega$ (the bandwidth of the pulse) 
transforms into a finite width in coordinate via the resonance
condition given by the electronic potentials. This collapse of the
measurement process assumes that the nuclear motion is frozen 
during the observation, i.e. the excitation is impulsive, and the
window operator $\Op{W}$ is independent of the state of the
system \cite{Jiri00}.

\section{Inducing the correlation dynamics by excitation with the pump
  pulse} 
\label{sec:pump}

\begin{figure}[bt]
  \centering
  \includegraphics[width=0.95\linewidth]{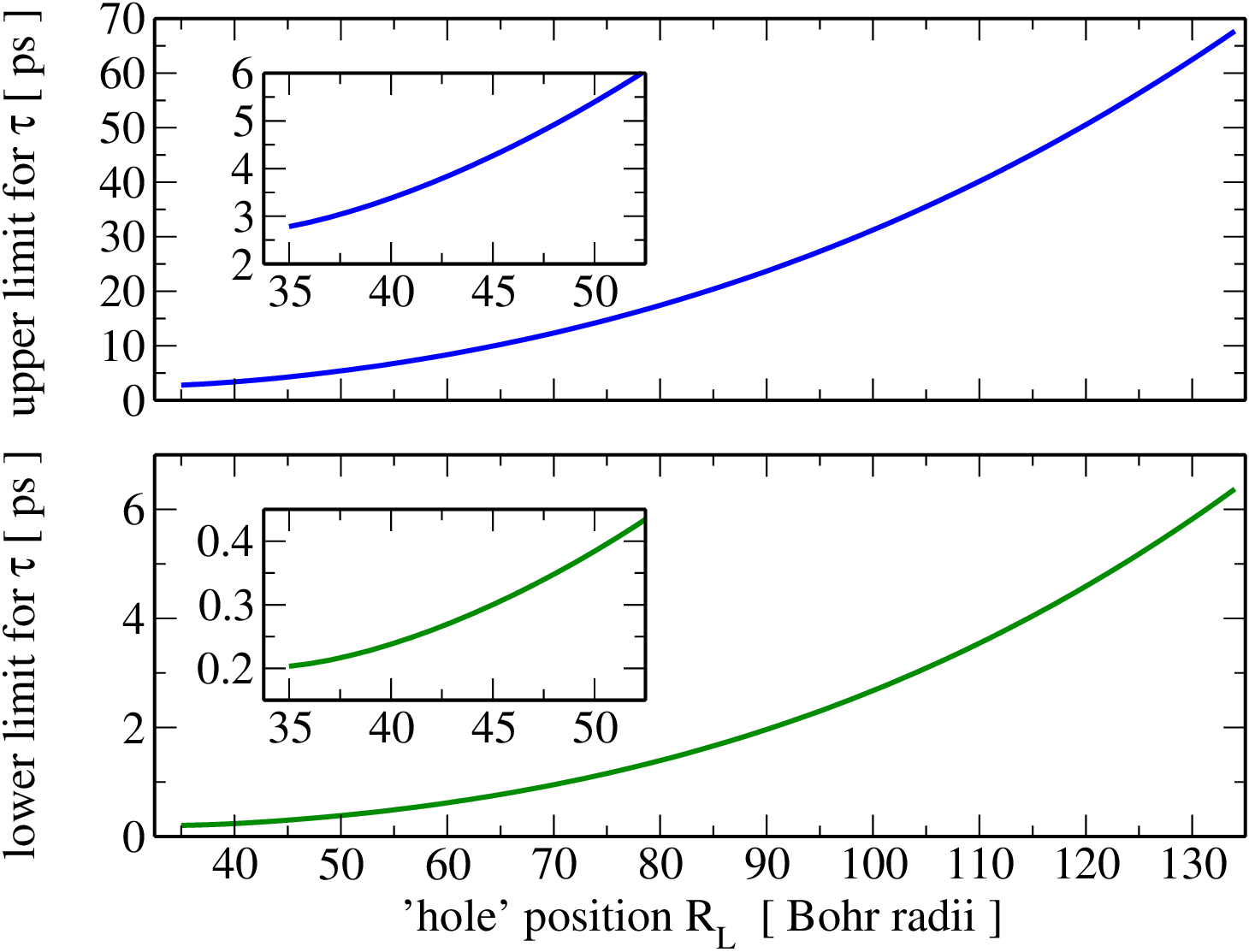}
  \caption{Upper and lower bounds on the transform-limited pulse
    duration $\tau$ according to Eqs.~\eqref{eq:ubound} and
    \eqref{eq:lbound}.} 
  \label{fig:tau}
\end{figure}

The dynamical pair correlations are initiated by an impulsive photoassociation
pulse. The resonance condition,  
$\Op \Delta(R_L)  = \Op V_\mathrm{exc} (R_L)- \Op V_g(R_L)-\hbar
\omega_L=0$, determines the central position 
of the dynamical 'hole', i.e. the Condon radius $R_L$.
The impulsive conditions on the pulse duration $\tau$ 
are determined by this point $R_L$. 
An estimate is based on the requirement that the energy gain due to
acceleration during the pulse, $\Delta E_\mathrm{kin}$,
is smaller than the energy bandwidth of the
pulse, $\hbar \Delta \omega = \hbar / \tau$. 
For a weak pulse such that excitation but no Rabi cycling occurs at
$R_L$ during the 
pulse, the energy gain is estimated by the acceleration due to the
difference potential $\Op \Delta(R)$, at the point of resonance, 
$R_L$. 
In a semi-classical picture, the gain in kinetic energy, is estimated
by integrating over the force, 
\[
\dot{P} = -\frac{\partial \Op{\Delta}}{\partial R}\bigg|_{R=R_L}\,,
\]
which yields the change in momentum,
$\Delta P = -\frac{\partial \Op{\Delta}}{\partial R}\big|_{R=R_L}\tau$.
Evaluating $\Delta E_\mathrm{kin} \,<\, \hbar\Delta\omega$ 
leads to an estimate for the upper limit of the pulse duration $\tau$,
\begin{equation}
\tau ~<~ \left(
  \frac{2 \mu\hbar}{\left(
      \frac{ \partial \Op \Delta}{\partial R}\big|_{R_L}
    \right)^2} \right)^{1/3}\,.
\end{equation}
For excitation in the asymptotic region where the difference potential
can be approximated to leading order by $-C_3/R^3$, this becomes
\begin{equation}
\tau_\mathrm{asy} ~<~ \left( \frac{2 \mu \hbar R_L^8}{9 C_3^2} \right)^{1/3}\,.
\label{eq:ubound}
\end{equation}
For the $0_g^-(P_{3/2})$ state of $^{87}$Rb one obtains $\tau<85\,$ps,
$300\,$ps, and $440\,$ps 
for $R_L=50\,$a$_0$, $78\,$a$_0$, and $90\,$a$_0$.
A lower limit to the duration of the pulse $\tau$ can be estimated due
to the requirement that the pulse 
does not excite the atomic transition, i.e. only pairs are excited. This means 
$\hbar\Delta\omega < |V_\mathrm{exc}(R_L)| \approx|C_3/R_L^3|$ or
\begin{equation}
\tau_\mathrm{asy} ~>~ \left|\frac{ \hbar R_L^3}{C_3}\right|\,.
\label{eq:lbound}
\end{equation}
For the $0_g^-(P_{3/2})$ state of $^{87}$Rb one obtains $\tau>385\,$fs,
$1.3\,$ps, and $2\,$ps 
for $R_L=50\,$a$_0$, $78\,$a$_0$, and $90\,$a$_0$.
The bounds on $\tau$ are plotted in Fig.~\ref{fig:tau}.
They indicate
how to determine the optimal transform-limited pulse duration of the 
photoassociation pulse as a compromise between impulsive
excitation and sharp spectroscopic features.
Note that the upper limit scales with $R_L^{2.67}$ while the lower limit
scales with $R_L^3$. The different scaling with $R_L$ implies
that there is a maximum radius $R^*_L$ for which a 'hole' can
be drilled in the pair correlation function. However, this maximum
radius amounts to $R^*_L\approx 160000\,$a$_0$ or $R^*_L\approx
8.5\,\mu$m with a corresponding $\tau$ of 10$\,$ms, so it does not
impose any practical limitations.
The region of interest for $R_L$ can be read off
Fig.~\ref{fig:scheme}: The minimum $R_L$ is due to the decrease in
scattering amplitude with shorter internuclear distance. The maximum
$R_L$ is determined in terms of the pulses that can feasibly be
produced in an experiment -- larger $R_L$ require longer delay
stages between pump and probe pulses and larger transform-limited
pulse durations, cf. lower panel of Fig.~\ref{fig:tau},
or smaller bandwidths, respectively.
In the following calculations, we choose a transform-limited pulse
duration of the photoassociation pulse of
$\tau=10\,$ps corresponding to an energy bandwidth of
1.5$\,$cm$^{-1}$. 

\begin{figure}[tb]
  \centering
  \includegraphics[width=0.95\linewidth]{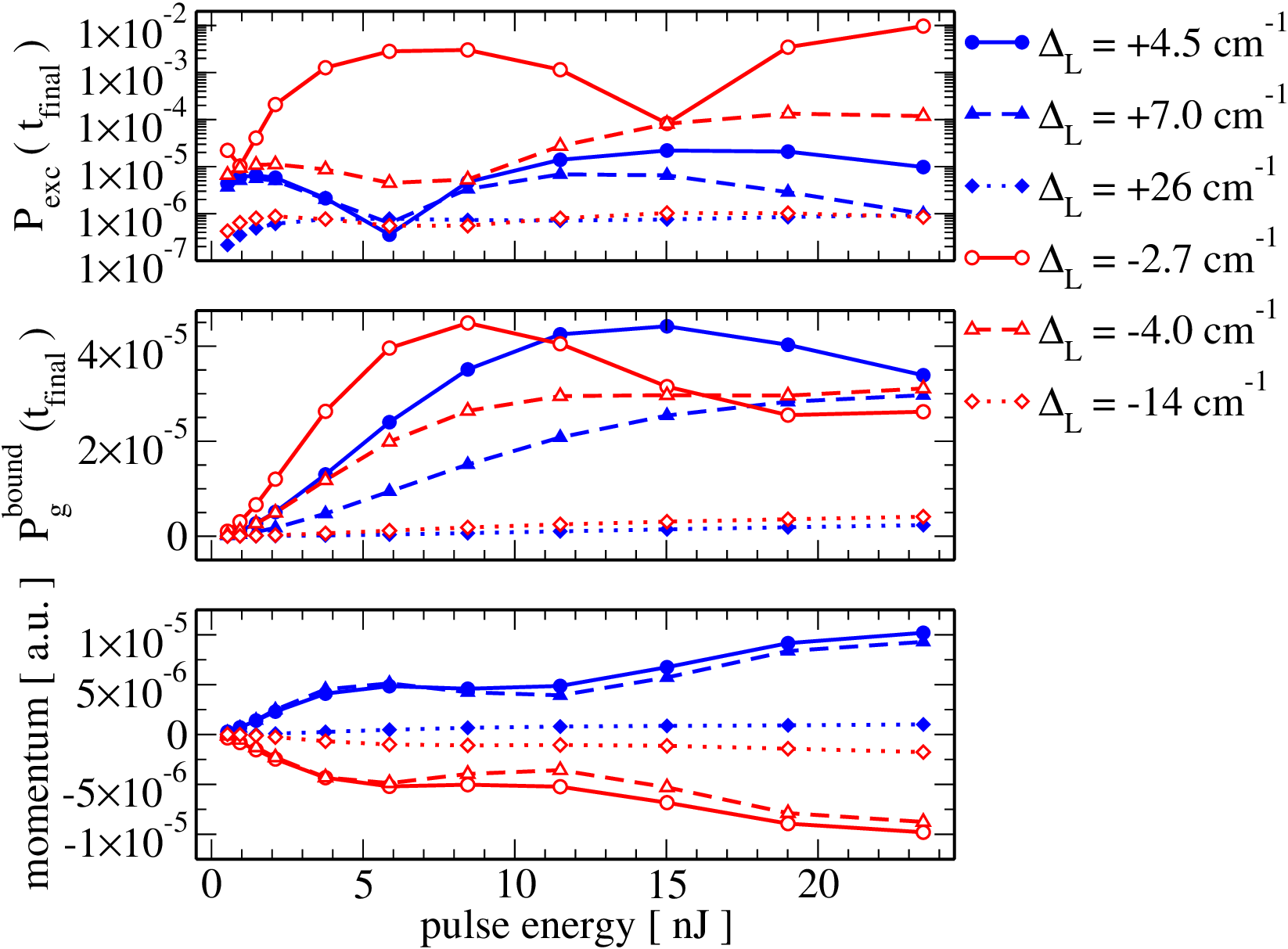}
  \caption{(Color online)
    The effect of the pump pulse: Population transfer to the excited
    state (top), formation of ground state molecules (middle) and
    momentum kick (bottom). Shown are the 
    excited state population (top), the population in bound levels of the
    electronic ground state (middle) and the ground state momentum
    (bottom) after the pump pulse is over as a function of the pulse
    energy. Different detunings are compared with corresponding
    Condon radii of 90$\,$a$_0$ (circles),  78$\,$a$_0$ (triangles),
    and 50$\,$a$_0$ (diamonds).
  }
  \label{fig:afterpulse}
\end{figure}

The photoassociation pump pulse transfers population to the
electronically excited state, cf. upper panel of
Fig.~\ref{fig:afterpulse}. 
 For small and medium detuning, Rabi
oscillations are observed in the population transfer to the excited
state, $P_\mathrm{exc}(t_\mathrm{final})$. The difference between the data for
$\Delta_\mathrm{L} = -2.7\,$cm$^{-1}$ and all other curves is explained by
excitation of the atomic resonance: the amplitude of a 10$\,$ps pulse
is still sufficiently large at zero detuning to excite population.  
The pump pulse does not only transfer
population to the excited state, it has also an effect on the ground state
wave function, creating the hole and transferring momentum, cf. lower
panel of Fig.~\ref{fig:afterpulse}. The  hole corresponds to
population of both bound levels and higher energy scattering states,
i.e. to the formation of  ground state molecules and hot atom pairs. 
The population of bound levels is shown in the middle panel of
Fig.~\ref{fig:afterpulse}. For small
and medium detuning, molecules are formed mostly in the last bound
level, $v=40$ (between 90\% and 99\%). 
The small amount of initial population near the
Condon radius is the limiting factor for the population of bound
levels at large detuning. In that case,
molecules in the last three levels are formed, with 55\% to 65\% in the
level $v=39$. These levels can be thought of as making up the
hole, i.e. they determine the hole dynamics.  
The bound population decreases at large pulse energies for small
detuning. This is due to power broadening, i.e. population residing
at the 'right' distance is transferred to continuum states 
rather than to bound levels. 

The momentum expectation value of the ground state wave function after
the pulse is shown in the bottom panel of
Fig.~\ref{fig:afterpulse}. For blue detuning, the hole is accelerated
toward larger distances, while for red detuning dynamics toward shorter
distances occur (as indicated in Fig.~\ref{fig:scheme}). Population enhancement
at short distance which is 
measured by the probe pulse can therefore only be expected for red
detuning. 

\begin{figure}[tb]
  \centering
  \includegraphics[width=0.9\linewidth]{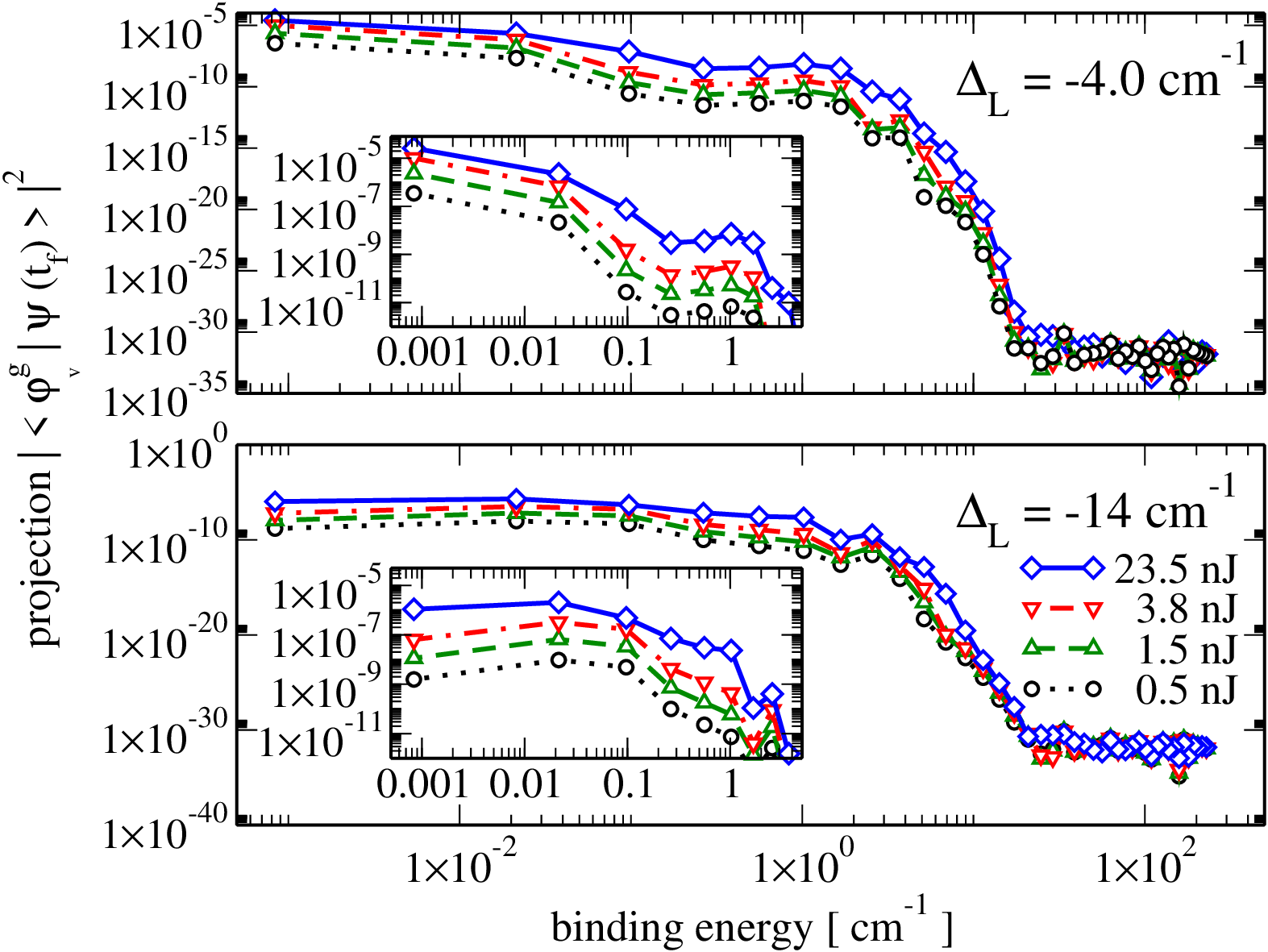}
  \caption{(Color online)
    Formation of ground state molecules due to the pump pulse. Shown
    is the projection of the final-time wave packet  onto the
    bound ground state vibrational levels for two red detunings.
  }
  \label{fig:vibdistg}
\end{figure}

While some bound levels are populated, population transfer occurs also into
 continuum states, and  the overall
 energy of the wave function is changed due to the interaction with
 the pump pulse. For red detuning, an increase in energy is always
 observed, while for blue detuning a decrease in energy occurs
 for medium and large detuning at large pulse energies.
The projection of the  wave packet  onto the
bound ground state vibrational levels after the pulse is shown in
Fig.~\ref{fig:vibdistg} for red detuning ($\Delta_\mathrm{L}=-4\,$cm$^{-1}$).
The vibrational distribution will be used to interpret the
time-dependent signals for weak-field excitation. 
Dynamical signatures of the levels with
binding energy of 1$\,$cm$^{-1}$ and less are expected. The
corresponding vibrational periods are between 59$\,$ps and
3$\,$ns. 

\section{Observation of the correlation dynamics by the probe pulse}
\label{sec:probe}

The probe pulse measures the two-atom amplitude at a
certain range of distances. The dynamics are unraveled by 
varying the delay time between pump and
probe pulse. Two-atom correlations may be detected by probing the
molecular part of the ground state wave function. Detection of
weakly bound ground state molecules has been developed in the context
of photoassociation experiments using cw excitation
\cite{FiorettiPRL98}. It is based on resonantly enhanced multi-photon
ionization.
In the following, two different detection schemes are explored.

(i) The dynamics of the hole can be probed when it arrives at short
internuclear distance in the inner region of the potential. This
implies the two-photon ionization scheme that has been utilized before
in photoassociation experiments to  detect molecules in the
lowest triplet state \cite{LozeilleEPJD06}. The resonant enhancement
is provided by the $^3\Sigma^+_\mathrm{g}(5s+4d)$ state,
cf. Fig.~\ref{fig:scheme}, i.e. the probe pulse has a  central frequency
that is  different from the pump pulse.
Assuming  ionization from the $^3\Sigma^+_\mathrm{g}(5s+4d)$ intermediate
state to be saturated, the detection is determined solely by the
Franck-Condon factors between the lowest triplet and the intermediate
states. This is reflected by the corresponding difference potential
that enters the window operator, Eq.~(\ref{eq:window}).

(ii) Alternatively, the hole can be probed at the position of its
creation. A possible REMPI scheme consists of overlapping a probe
pulse that has the same central frequency as the pump pulse with
another pulse ionizing from the $0_\mathrm{g}^-(5s+5p_{3/2})$  excited
state. Such an ionization scheme has been utilized before in
femtosecond photoassociation experiments to detect excited state
molecules \cite{SalzmannPRL08}. Assuming the detection to be
determined by the probe pulse, the difference potential of the
$0_\mathrm{g}^-(5s+5p_{3/2})$ and the lowest triplet states enters the window
operator.

In the second detection scheme, one needs to ensure that the excited
state wave packet that is created by the pump pulse does not interfere
with the density that is excited by the probe pulse. This can achieved
by sending another ionization pulse simultaneously with 
the pump pulse such that any initial excited state density is
immediately eliminated from the sample. The overall pump-probe scheme
consists therefore of two time-delayed pulse pairs, pump + ionization
pulse followed by probe + ionization pulse. Since all pulses just
address wave packet density in a certain range of internuclear
distances, no phase relation of the four pulses is required.

\subsection{Probing the hole at short and intermediate distances}
\label{subsec:probe1}

Since the excitation by the pump pulse leads to a redistribution of
the ground state probability, the hole contains a part
corresponding to bound molecules and a part corresponding to hot
atoms. Overall, the hole is accelerated toward short distances for red
detuning of the pump pulse. The signal is therefore expected to
consist of two parts: At early times, a large peak reflects the
arrival of both molecular and hot atomic density in the probe
window. After the first reflection at the inner
turning point of the potential, the hot atomic density leaves the
short-range distances. Only the molecular part continues to
move within the range of the potential. Small
oscillations reflect the motion of the molecular part at later
observation times.

\begin{figure}[tb]
  \centering
  \includegraphics[width=0.95\linewidth]{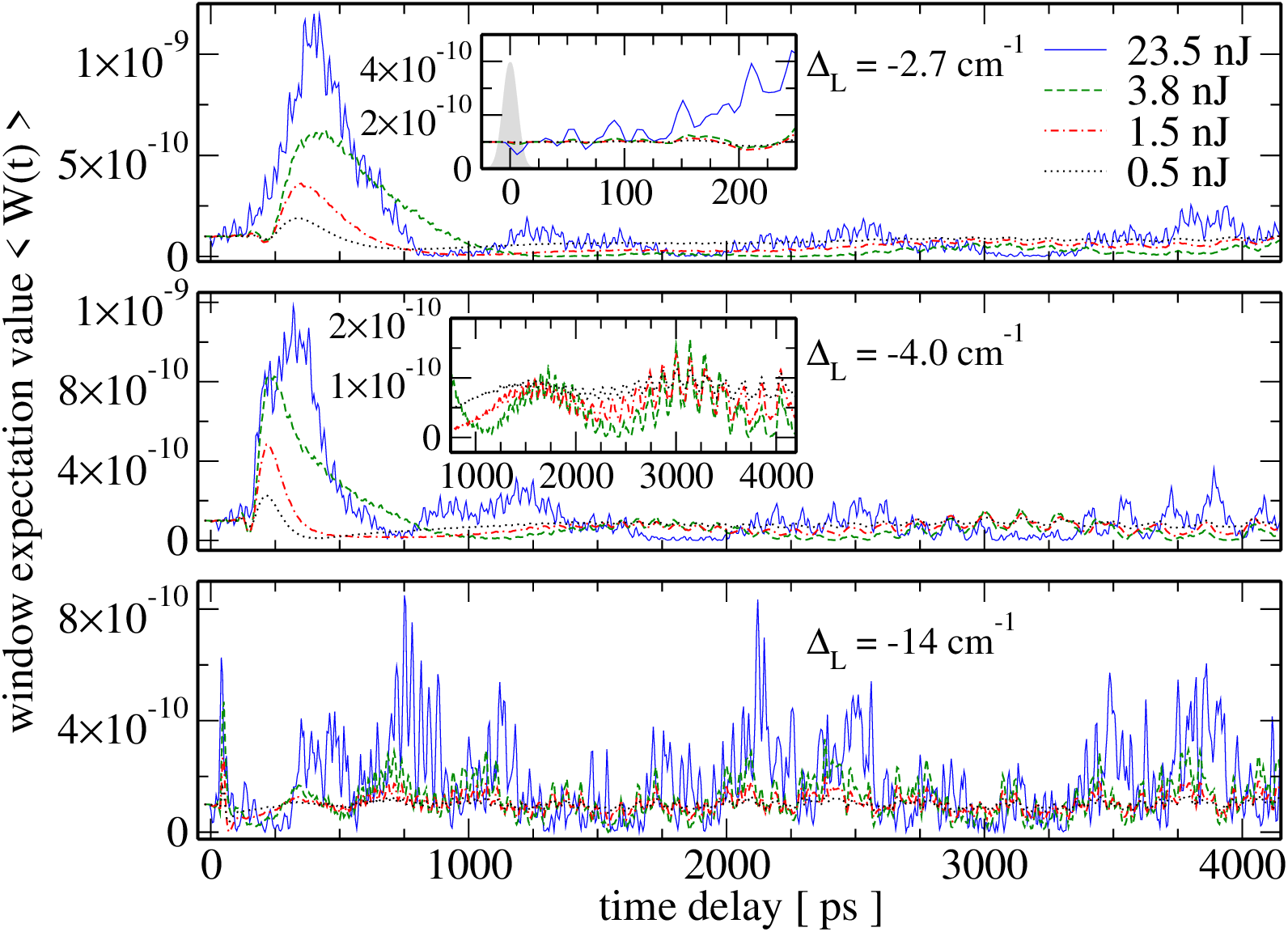}
  \caption{(Color online)
    Probing the dynamical hole near the inner turning point: The
    time-dependent expectation value of the window operator is plotted 
    for different energies and detunings of the pump pulse and a probe wavelength of
    680$\,$nm (window localized around $R=10\,$a$_0$).
  }
  \label{fig:window1}
\end{figure}

\begin{figure}[tb]
  \centering
  \includegraphics[width=0.95\linewidth]{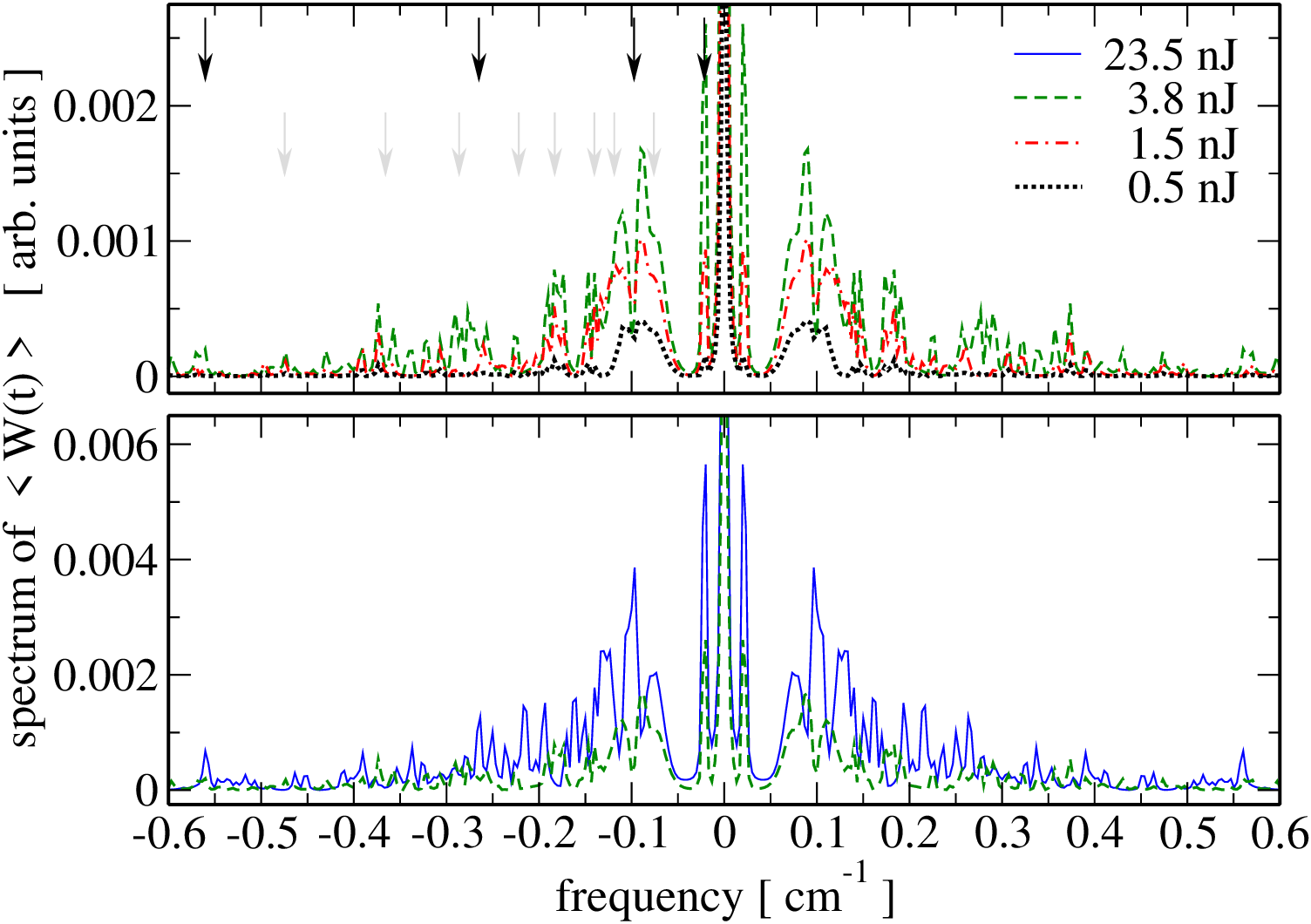}
  \caption{(Color online)
    Spectrum of the time-dependent expectation value of the window
    operator (middle panel of Fig.~\ref{fig:window1}, i.e.
    $\Delta_\mathrm{L}=-4\,$cm$^{-1}$ and a probe window around
    $R=10\,$a$_0$).
    The black arrows indicate the position of ground state
    vibrational levels, the grey arrows some of the  beat
    frequencies between different vibrational levels.
    }
  \label{fig:specwindow1}
\end{figure}

\begin{figure}[tb]
  \centering
  \includegraphics[width=0.95\linewidth]{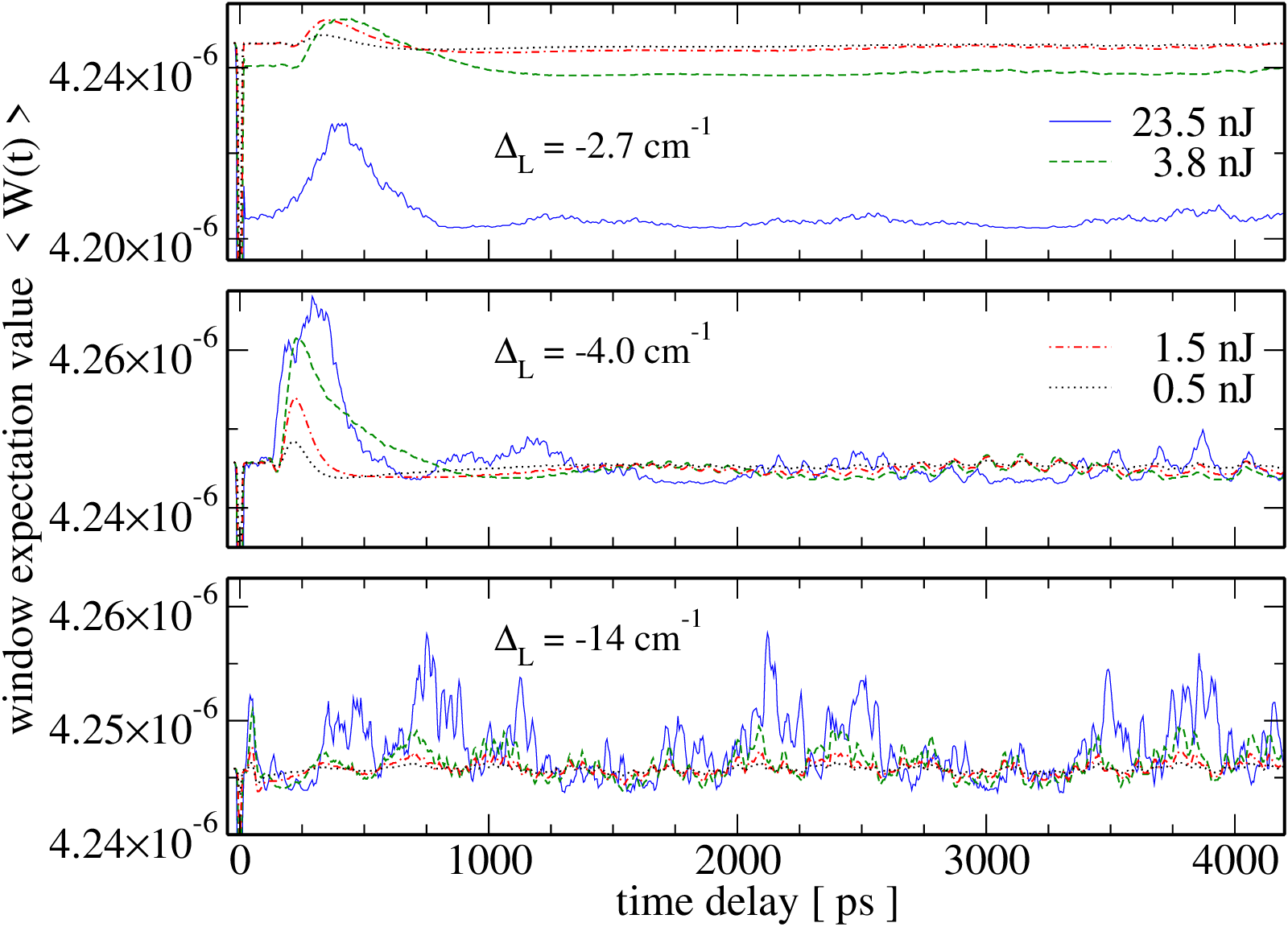}
  \caption{(Color online)
    Probing the dynamical hole at intermediate distances: The
    time-dependent expectation value of the window operator is plotted
    for different energies and detunings of the pump pulse and a probe wavelength of
    518.5$\,$nm (window localized around $R=24\,$a$_0$).    
    }
  \label{fig:window2}
\end{figure}

\begin{figure}[tb]
  \centering
  \includegraphics[width=0.95\linewidth]{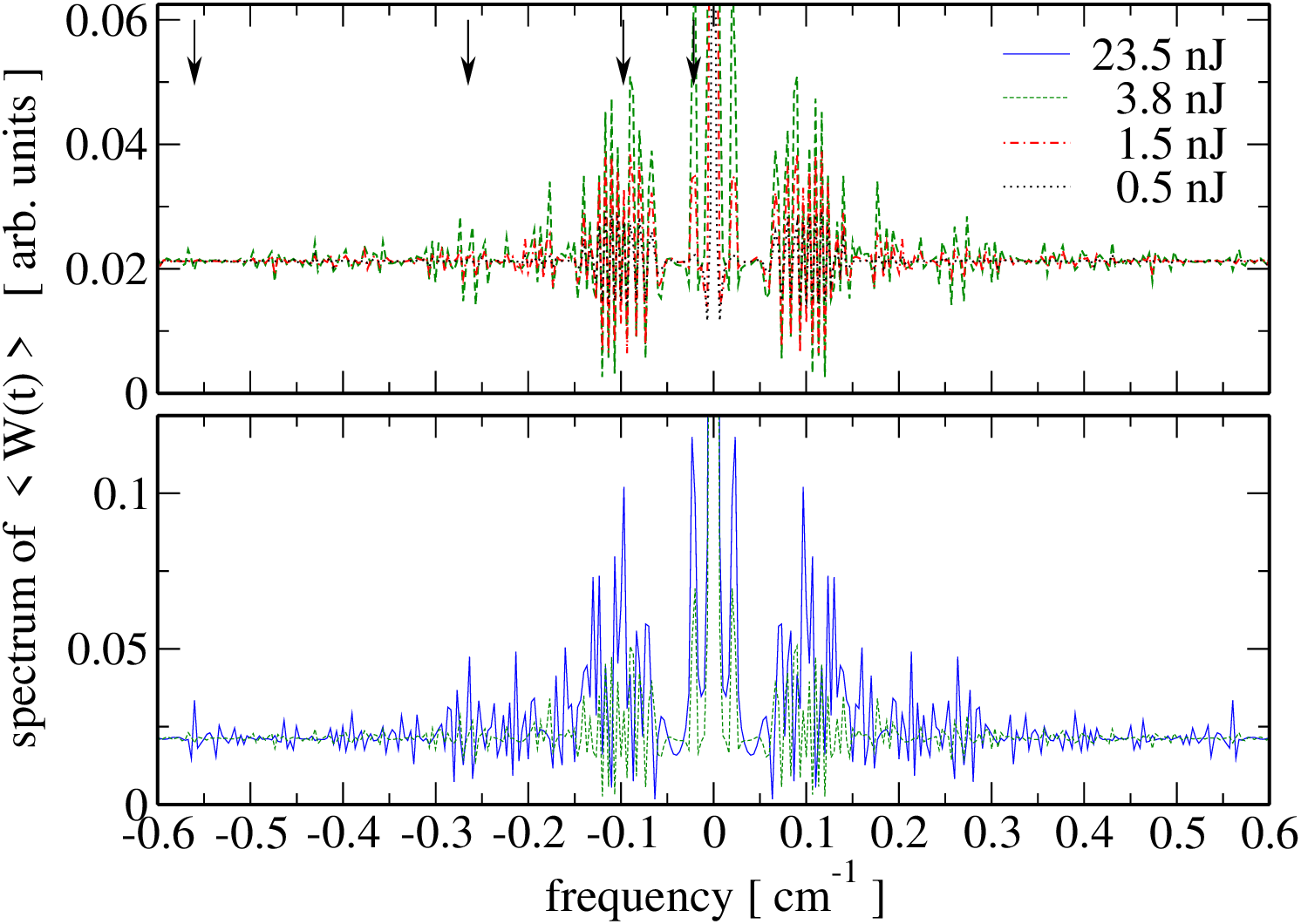}
  \caption{(Color online)
    Spectrum of the time-dependent expectation value of the window
    operator (middle panel of Fig.~\ref{fig:window2}, i.e.
    $\Delta_\mathrm{L}=-4\,$cm$^{-1}$ and a probe window around  $R=24\,$a$_0$). 
    The solid arrows indicate the position of ground state
    vibrational levels.
  }
  \label{fig:specwindow2}
\end{figure}

This behavior is indeed observed in the time-dependent
 expectation value of the window operator modeling the
probe pulse absorption in Figs.~\ref{fig:window1} and
\ref{fig:window2}. $\langle \Op{W}(t)\rangle$
is shown as a function of the pump-probe delay for probe
wavelengths of 680$\,$nm with 20$\,$nm bandwidth and of 518.5$\,$nm with
1$\,$nm bandwidth, respectively. The spectra of these signals
are obtained by filter-diagonalization
\cite{MandelTaylor97,MandelTaylor98}, a method allowing to accurately
extract frequencies from just a few oscillation periods. The spectra of the
probe signals of the middle panels of Figs.~\ref{fig:window1} and
\ref{fig:window2} are shown in Figs.~\ref{fig:specwindow1} and
\ref{fig:specwindow2}. 

For the probe measurement at $R=10\,$a$_0$
($\omega_P=680\,$nm), cf.  Fig.~\ref{fig:window1}, 
an enhancement of the signal by a factor up to 10
is observed at short delay times. The enhancement is larger for
smaller  pump pulse  detuning since more population is excited
creating a larger hole, cf. Fig.~\ref{fig:afterpulse}.  
It is due to the accelerated
population which contains both bound and continuum parts. At longer
delay times, only the bound part of the wave packet returns to
$R=10\,$a$_0$. This part of the dynamics is characterized by a large
oscillation with overtones. It can be analyzed in terms of the
vibrational periods of the molecular levels making up the bound part
of the wave packet. The oscillation period corresponds to the
vibrational period of the last, respectively last but one, level,
cf. Fig.~\ref{fig:vibdistg}. The frequency of the long-term oscillations
increases with  pump pulse  detuning. This may be rationalized in terms of
the vibrational distributions, cf. Fig.~\ref{fig:vibdistg}.  

The overtones represent a beating between the different
vibrational levels making up the hole. This becomes evident
in Fig.~\ref{fig:specwindow1} where the spectrum of the probe signal
is plotted for the detuning $\Delta_\mathrm{L}=-4\,$cm$^{-1}$ and
different energies of the pump pulse.
The solid arrows indicate the spectral positions of the ground state
vibrational levels, except for the last bound level which is too close
to zero to be visible ($E_\mathrm{bind}^\mathrm{last}=8.4\times 10^{-4}$cm$^{-1}$ or
25 MHz). Of the remaining levels, the second to last level
at $E_\mathrm{bind}^\mathrm{last-1}=2.1\times 10^{-2}$cm$^{-1}$
draws the highest peak for weak pump  pulses
in accordance with the vibrational distribution, cf.
lower panel of Fig.~\ref{fig:vibdistg}. In the weak-field regime,
upper panel of Fig.~\ref{fig:specwindow1}, the peaks in the spectrum
appear at the same positions for different pump pulse energies.
Increasing the pulse energy leads to more population in the lower
lying vibrational levels and hence additional features,
i.e. beat frequencies, appear in the
spectrum. The strong-field and weak-field regime are compared in the
lower panel of Fig.~\ref{fig:specwindow1}. At a pump pulse energy of
23.5$\,$nJ, Rabi cycling occurs and significantly more momentum is
transferred to the ground state wave packet than for 3.8$\,$nJ,
cf. lower panel of Fig.~\ref{fig:afterpulse}. This results in a shift
of the spectral peak positions, cf. the solid blue and dashed green
lines in the lower panel of Fig.~\ref{fig:afterpulse}: For strong
field, so much energy is pumped into the system that the dynamics is
not easily unraveled in terms of a decomposition into field-free
vibrational periods.


The overall probe signal is much larger for a 
measurement at $R=24\,$a$_0$ ($\omega_P=518.5\,$nm ) than for one at
$R=10\,$a$_0$ ($\omega_P=680\,$nm). However, the \textit{relative}
enhancement is  significantly smaller, cf. Fig.~\ref{fig:window2}.
Since the measurement occurs at larger distances, there is
ground state amplitude initially within the probe window,
cf. Fig.~\ref{fig:scheme}, leading to a
strong background. The pump pulse cycles population which
shows up as a dip in the probe signal around zero delay,
cf. Fig.~\ref{fig:window2}. Due to the shape of the potentials,
the probe window is much broader for $\omega_P=518.5\,$nm than for
$\omega_P=680\,$nm for the same probe spectral bandwidth.
The oscillations and overtones at large delays are therefore less
clearly resolved for $\omega_P=518.5\,$nm. Similarly to
Fig.~\ref{fig:window1}, larger pump pulse detuning leads to faster
oscillations and larger pump pulse energy to a stronger signal in
Fig.~\ref{fig:window2}. 

The probe spectra for  a measurement at $R=24\,$a$_0$
($\omega_P=518.5\,$nm ) and a pump pulse detuning of
$\Delta_\mathrm{L}=-4\,$cm$^{-1}$ are analyzed in
Fig.~\ref{fig:specwindow2}. The signal background leads to a non-zero
offset of the spectral baseline. The positions of the peaks are again
compared to those of the vibrational levels, indicated by  black
arrows. As expected they agree very well, with the beating between
different vibrational levels leading to a splitting of the peaks. For a
strong pump pulse, similarly to Fig.~\ref{fig:specwindow1} a shift of the
peak positions is observed compared to smaller pump pulse energies.
The similarities in Figs.~\ref{fig:specwindow1} and
\ref{fig:specwindow2}, i.e. peak positions and shifts,
reflect the underlying wave packet dynamics which is
completely determined by the (identical) pump pulse excitation. The differences
such as the spectral baseline and specific peak shapes
are attributed to the differing ways of measuring the dynamics.

The third window operator shown in Fig.~\ref{fig:scheme} that corresponds
to a probe wavelength of 516.5$\,$nm and to a probe bandwidth of
1$\,$nm does not show any time-dependence  (data not shown).
The population over the broad range of distances covered by this
window does not sufficiently vary. 

In conclusion, probing the correlation dynamics by ionization via the 
$^3\Sigma^+_\mathrm{g}(5s+4d)$ state allows for identifying the
positions of the last bound levels of the ground state potential,
provided time delays of a few nanoseconds can be realized. Probing the
dynamics at very short distance, $R=10\,$a$_0$, yields a signal that 
clearly oscillates as a function of the pump-probe time delay. The
strength of the probe signal is, however, somewhat discouraging. While
a larger signal can be achieved by probing the
dynamics at intermediate distance, $R=24\,$a$_0$, the contrast of the
oscillations is diminished due to the broader probe window.

\subsection{Probing the hole at the position of its creation} 
\label{subsec:probe2}

\begin{figure}[tb]
  \centering
  \includegraphics[width=0.95\linewidth]{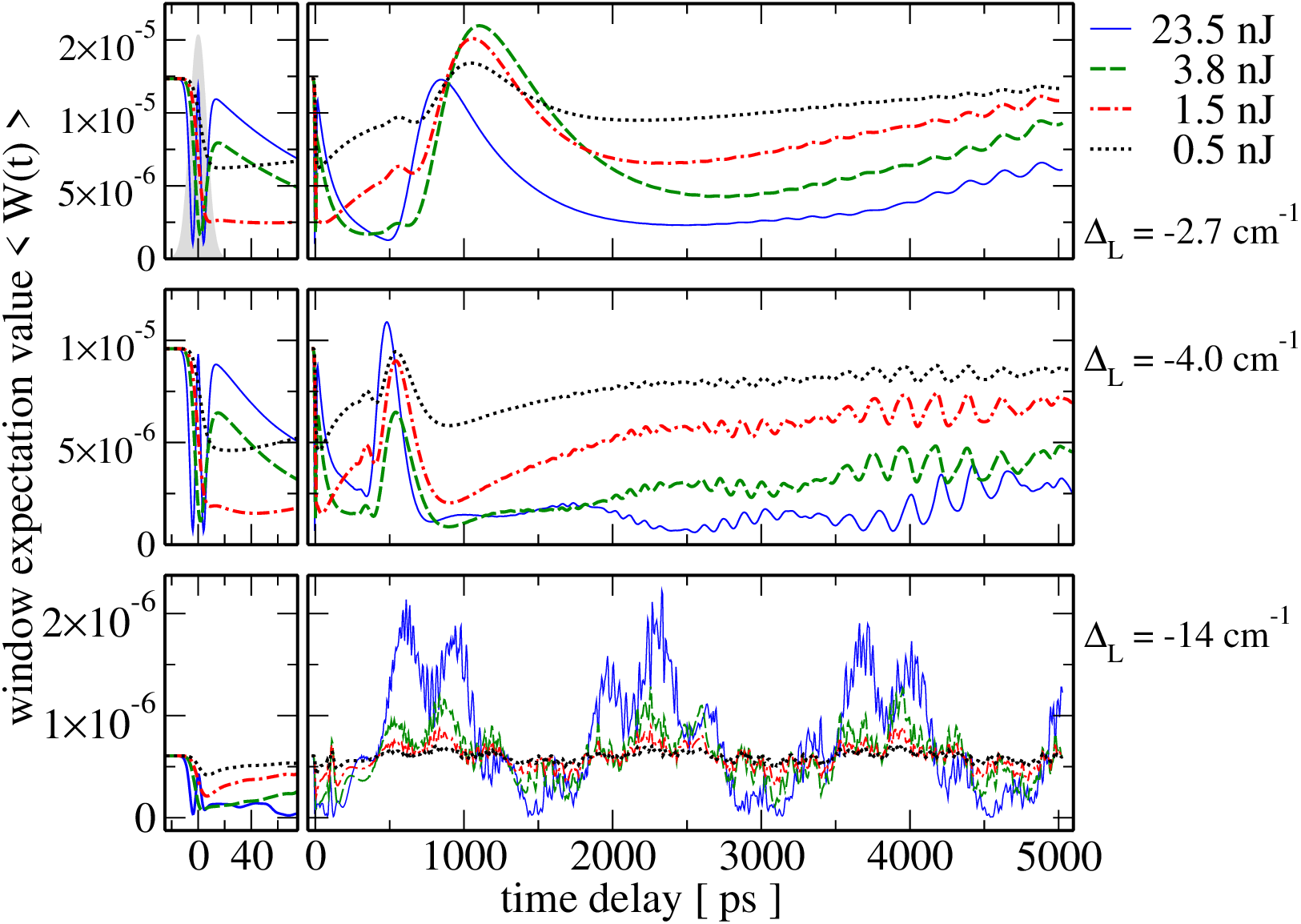}
  \caption{(Color online)
    Probing the dynamical hole at the position of its creation: The
    time-dependent expectation value of the window operator is shown
    for different pump pulse energies and a probe wavelength of
    780.5$\,$nm (window localized around $R=90\,$a$_0$ for
    $\Delta_\mathrm{L}=-2.7\,$cm$^{-1}$,
    $R=78\,$a$_0$ for $\Delta_\mathrm{L}=-4.0\,$cm$^{-1}$, and
    $R=49\,$a$_0$ for $\Delta_\mathrm{L}=-14\,$cm$^{-1}$).}
  \label{fig:window3}
\end{figure}

A similar time-dependence of the probe signal is expected for 
probing the correlation dynamics at the position where the hole
is created and at short internuclear distances, i.e. a large
enhancement at short times due to both the hot atomic and the
molecular components of the hole and smaller oscillations at longer
times that can be attributed to the molecular
components. Additionally, when probing  at the position of the hole
creation and with pump and probe pulses that overlap in time,
around zero delay, a strong dip of the probe signal should indicate
transient population transfer to the excited state. 
These expectations are confirmed by inspecting the probe signals shown
in Fig.~\ref{fig:window3}. Moreover, a large signal strength and a strong
relative enhancement or, respectively, depletion are observed.

\begin{figure}[tb]
  \centering
  \includegraphics[width=0.95\linewidth]{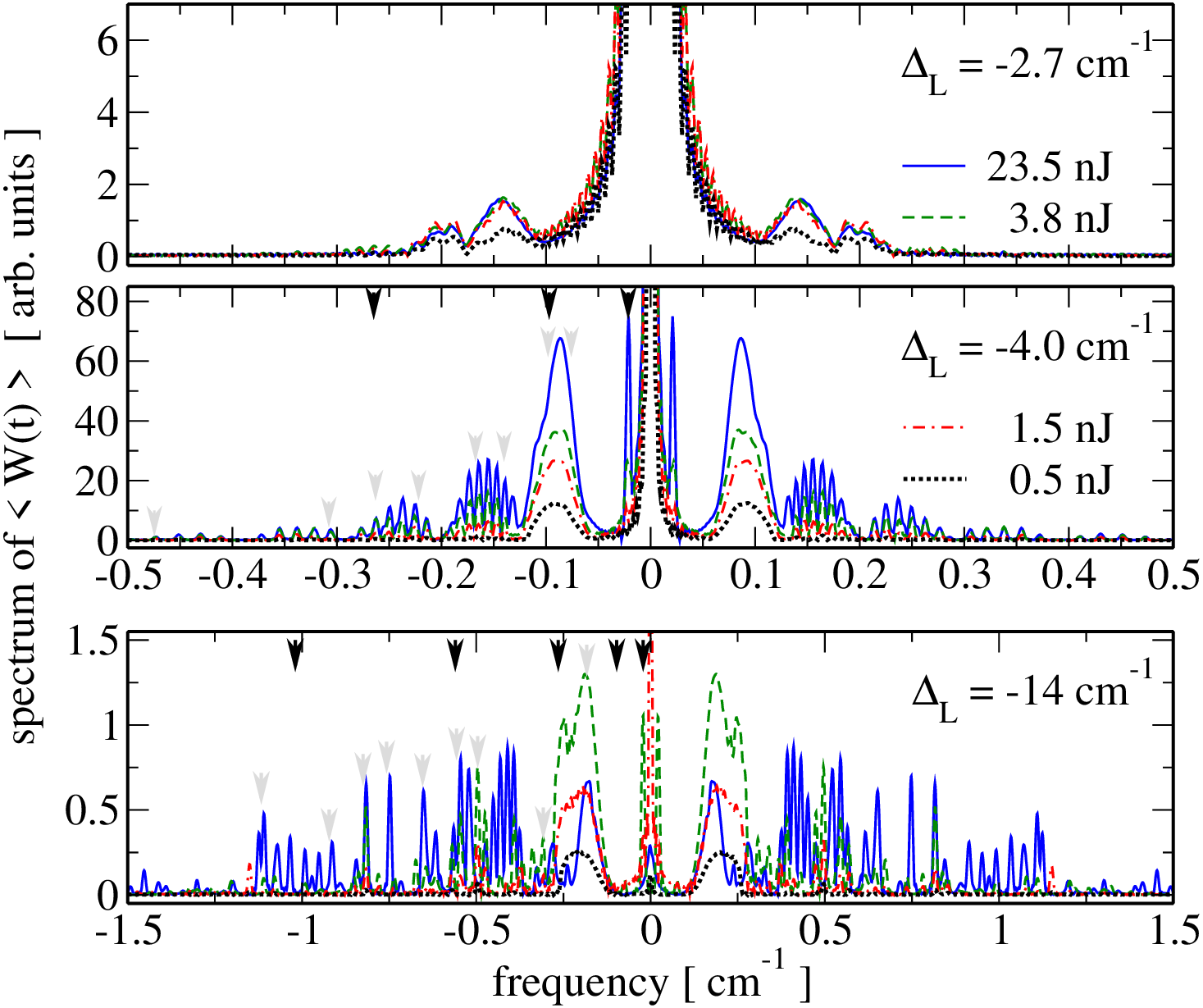}
  \caption{(Color online)
    Spectrum of the time-dependent expectation value of the window
    operator for $\Delta_\mathrm{L}=-2.7\,$cm$^{-1}$ and a probe
    window around  $R=90\,$a$_0$,
    $\Delta_\mathrm{L}=-4\,$cm$^{-1}$ and a probe window around
    $R=78\,$a$_0$ (upper panel) and for
    $\Delta_\mathrm{L}=-14\,$cm$^{-1}$ and a
    probe window around  $R=49\,$a$_0$ (lower panel).
    The black arrows indicate the position of ground state
    vibrational levels, the gray arrows some of the  beat
    frequencies between different vibrational levels.
  }
  \label{fig:specwindow3}
\end{figure}

The motion of the ground state density toward the inner turning point
followed by the refilling of the hole is reflected in
the peak of the signal which occurs, depending on the pump pulse energy,
between 840$\,$ps and 1100$\,$ps for
$\Delta_\mathrm{L}=-2.7\,$cm$^{-1}$,  between 480$\,$ps and 540$\,$ps for
$\Delta_\mathrm{L}=-4.0\,$cm$^{-1}$, and  between 100$\,$ps and 110$\,$ps for
$\Delta_\mathrm{L}=-14\,$cm$^{-1}$. The refilling of the hole corresponds
to the 'recovery of the bleach' known from molecular pump-probe spectroscopy.
Unlike those examples from gas-phase or condensed-phase dynamics, the
signal in our case is not caused by vibronic motion of molecules but
rather by the two-body correlations between  atoms that are present in
ultracold gases.

The highest pump pulse energy
always yields the fastest refilling. This is attributed to the
momentum transfer which increases with pulse energy,
cf. Fig.~\ref{fig:afterpulse}. After the initial refilling, another
depletion is observed followed by a slow recovery with
small oscillatory modulations of the
signal. As in Section~\ref{subsec:probe1}, a
larger detuning of the pump pulse leads to a higher frequency of these
oscillations. The amplitudes of the oscillations can be enhanced by
increasing the pump pulse energy, in particular for large detunings. 

The strength of the probe signal without any pump pulse is given
by the probability density amplitude at the Condon radius. For larger
detuning, the Condon radius and hence the amplitude decrease,
cf. Fig.~\ref{fig:scheme}. This determines the strength of the
background signal from which the 'recovery of the bleach' is
measured. It is decreased by an order of magnitude when
increasing the pump pulse detuning from
$\Delta_\mathrm{L}=-4.0\,$cm$^{-1}$ to
$\Delta_\mathrm{L}=-14\,$cm$^{-1}$ (cf. initial values in
Fig.~\ref{fig:window3}). 

It might seem a little surprising at first sight, that an enhancement
of the signal \textit{above} its background value is observed for
strong field and large detuning, cf. blue solid curve in the lowest
panel of Fig.~\ref{fig:window3}. This implies that, at certain times,
corresponding to pump-probe delays of 
e.g. 610$\,$ps or  2300$\,$ps, more probability density resides within
the probe window than is initially there. One would rather expect
that, after the initial refilling of the hole, when the  hot atomic
component of the ground state wave packet has left the short
internuclear distances of the probe window, the signal could not
be completely recovered anymore. This picture holds, however, only for
weak pump pulses. For a pump pulse energy of 23$\,$nJ, Rabi cycling
occurs, cf. upper panel of Fig.~\ref{fig:afterpulse}. It is
accompanied by power broadening. A broadened pulse spectrum
corresponds to addressing initial ground state density in a larger range of
internuclear distances. Therefore the  23$\,$nJ-pump pulse at a
detuning of $\Delta_\mathrm{L}=-14\,$cm$^{-1}$ 
excites not only the peak in the initial ground state wave function
around $50\,$a$_0$ but also part of the two neighboring peaks,
cf. Fig.~\ref{fig:scheme}. Since some of this population is cycled
back to the ground state, it may contribute to the bound components of
the hole and show up in the long-term oscillations of the probe
signal. 

The spectra of the probe signals are examined in
Fig.~\ref{fig:specwindow3}. The highest peaks occur at the position of
the last bound levels of the ground state potential that are indicated
by the black arrows. The peak around
zero is due to the last level which is not resolved on the shown
frequency scales  ($E_\mathrm{bind}^\mathrm{last}=8.4\times
10^{-4}$cm$^{-1}$ or 25 MHz).  It is off the top of the figure in the
upper two panels  for the smaller pump pulse detunings
but reduced in the lower panel  for the large pump pulse
detuning. This is in
accordance with Fig.~\ref{fig:vibdistg} where for the larger detuning
a smaller population of the last bound level is observed. 
The smaller peaks occurring at higher frequencies are attributed to
beat frequencies between different vibrational levels, some of which are
indicated by the gray arrows. Increasing the pump pulse energy leads
to larger peaks and a finer resolution of the spectral features, in
particular for the larger pump pulse detuning. A larger detuning of
the pump pulse implies larger components in the ground state wave
packet with binding energies above
$0.1\,$cm$^{-1}$, cf.  Fig.~\ref{fig:vibdistg}. This is reflected by
peaks occurring at higher frequencies in the spectrum (note the
different scales in the upper and lower panels of
Fig.~\ref{fig:specwindow3}). For a very small pump pulse detuning,
$\Delta_\mathrm{L}=-2.7\,$cm$^{-1}$ and probing at fairly long range,
$R=90\,$a$_0$, the dynamics are rather slow. An observation time of
5$\,$ns is then not sufficient to resolve the spectral features of the
probe signal.






\section{Conclusions}
\label{sec:concl}

A hole in the pair density of an ultracold gas is the consequence of
sudden unitary 
population transfer from the ground to the excited electronic state.
It represents a non-stationary state of the two-atom scattering whose
evolution in time can be monitored by a suitably chosen probe
pulse. Such a pump-probe spectroscopy of pair correlations can be
implemented by slight modification of existing experimental
setups
\cite{SalzmannPRL08,WeiseJPB09,MullinsPRA09,MerliMyPRA09,McCabePRA09}.  
It can be applied to Bose-Einstein condensates as well as thermal
ultracold gases. This corresponds to studying the
dynamics of a pure state vs those of an incoherent ensemble for
the timescales probed in such an experiment. The main effect of
thermal averaging is the contribution of higher partial waves. This is 
particularly prominent in the presence of resonances \cite{MyPRL09}. 
If a resonance affects different scattering channels differently, such
as shape resonances in either the singlet or the triplet channel of
rubidium, pump-probe spectroscopy can be used to map out the coupled
channels pair wavefunction despite the finite lifetime of the
resonance \cite{MyPRL09}. 

In the present work, we have studied the influence of the pulse
parameters on the hole dynamics for a pure initial state. 
Creation of a hole can be induced by a 
short pulse which is faster than the timescale of the nuclear dynamics. 
A practical requirement is that this pulse should be transparent at
the atomic transition.  
In that case, only pair correlations are dynamically modified. The pulse
has to be impulsive compared to the timescale of relative motion on
the electronic ground state. For ultracold rubidium, the pulse
duration should be a few picoseconds or, respectively, 
the spectral bandwidth a few wavenumbers. The same process could be
applied to other species with the timescale modified to maintain the
impulsive condition. 

The hole represents a non-stationary superposition of many scattering
states and a few of the last bound levels. The shape of the hole in phase space,
i.e. its momentum and the position, are controlled by
the pulse intensity, chirp and detuning from the atomic line. As a result there is
significant experimental flexibility to interrogate specific
properties of the pair correlation. 
The timescale of the dynamics of the hole can be estimated by
comparing the acceleration due to the difference potential and the
energy associated to the bandwidth of the pulse.  
An upper bound  of the transform-limited pulse duration is related to
the gain in kinetic energy due to the gradient of the difference
potential. A lower bound is imposed by the restriction not to excite
the atomic resonance. For all 'hole' positions of interest, this
defines a window of possible transform-limited pulse durations which
for rubidium are of the order of a few picoseconds to tens of picoseconds.

To monitor the dynamical evolution of the hole, a second weak pulse,
delayed in time from the pump pulse, is applied as a probe. 
Typically the number of pairs is small which means a high sensitivity is required.
In this study a two-photon REMPI scheme is suggested. The probe
pulse transfers amplitude to an intermediate state which then is
ionized leading to a high detection efficiency. 
We have modeled the absorption of the probe pulse by
time-dependent perturbation theory applied to the first step.
This assumes saturation of the second step in the REMPI scheme. 
The characteristics of the probe pulse duration, chirp and central frequency 
determines the properties of the measurement. 

Probing the two-body correlation dynamics yields a signal with clear
dynamical features. If one is able to provide for pump-probe delays of
a few nanoseconds, spectral features on a scale of less than $1\,$cm$^{-1}$
can be resolved. Since pump and probe just address probability density
at certain internuclear distances, no phase relation between the
pulses is required, and long time delays between the pump and probe
pulse can be realized by optical delay stages. 

The full power of coherent control can be employed to modify the
hole. The simplest modification makes use of chirped pulses. Chirping
the pump pulse changes the shape of the hole, while chirping the probe
pulse generates a measurement of position and momentum in phase space
with accuracy limited by the uncertainty relation.

This pump-probe spectroscopy of the pair correlation dynamics can be
combined with static external field control of the initial pair
density. Specifically, tuning a magnetic field close to a Feshbach
resonance enhances the pair density at short and intermediate
distances \cite{PellegriniPRL08}. Obviously, this will lead to a
larger probe signal strength. It remains an interesting open problem
to study the effect of the Feshbach resonance on the dynamical
properties.  


\begin{acknowledgments}
We would like to thank Fran\c{c}oise Masnou-Seeuws and Pascal Naidon
for many fruitful discussions. 
We gratefully acknowledge financial support  from 
the Deutsche Forschungsgemeinschaft.
The Fritz Haber
Center is supported by the Minerva Gesellschaft f\"{u}r die Forschung
GmbH M\"{u}nchen, Germany. 
\end{acknowledgments}


\end{document}